\documentclass[aps,twocolumn,prd,preprintnumbers,amsmath,amssymb,superscriptaddress,nofootinbib]{revtex4-1}
\pdfoutput=1

\usepackage[export]{adjustbox}

 \usepackage{color}
 \usepackage{epsfig}
 \usepackage{ulem}

\definecolor{White}{rgb}{1,1,1}
\definecolor{Red}{rgb}{1,0.1,0}
\definecolor{LightYellow}{rgb}{1,1,.875}
\definecolor{SteelBlue}{rgb}{.273,.508,.703}
\definecolor{navy}{rgb}{0,0,.5}
\definecolor{LightCyan}{rgb}{.875,1,1}
\definecolor{DarkRed}{rgb}{.543,0,0}
\definecolor{HotPink}{rgb}{1,.41,.70}
\definecolor{ForestGreen}{rgb}{.13,.54,.13}
\definecolor{OliveDrab}{rgb}{.42,.55,.14}
\definecolor{MediumBlue}{rgb}{0,0,.80}
\definecolor{RoyalBlue}{rgb}{.25,.41,.88}
\definecolor{DeepSkyBlue}{rgb}{0,.746,1}
\definecolor{Brown}{rgb}{0.545,0.271,0.074}
\definecolor{Purple}{rgb}{0.637,0.285,0.641}

\begin{document}

\title{Axion-philic cosmological moduli}

\author{Kwang Sik Jeong} 
\email{ksjeong@pusan.ac.kr}
\affiliation{Department of Physics,   Pusan National University,   Busan 46241,   Korea}
\author{Wan Il Park} 
\email{wipark@jbnu.ac.kr}
\affiliation{Division of Science Education and Institute of Fusion Science, 
Jeon-buk National University,   Jeonju,   Jeonbuk 54896,   Korea} 
\preprint{PNUTP-21-A14}

\begin{abstract} 

We show that string moduli have axion-philic nature owing to the model-insensitive derivative interactions
arising from the K\"ahler potential. 
The decay of a modulus into stringy axions occurs without suppression by the mass of final states.
Interestingly,   it turns out to hold in general  not only for the scalar partner of the stringy axion but also for any
other moduli. 
The decay into (pseudo) Nambu-Goldstone bosons (NGBs) also avoids such mass suppression 
if the modulus is lighter than or similar in mass to the scalar partner of the NGB.    
Such axion-philic nature makes string moduli a natural source of an observable amount of dark radiation
in string compactifications involving ultralight stringy axions,  and possibly in extensions of the Standard
Model that include  a cosmologically stable NGB such as the QCD axion.  
In the latter case,   the fermionic superpartner of the NGB can also contribute to 
the dark matter as a feebly interacting massive particle.

\end{abstract}

\maketitle

\section{Introduction}

String compactification involves many moduli with various masses.
Their Planck-suppressed couplings make them long-lived,   and typically
the lightest one most relevant to cosmology.
Displaced from the post-inflationary potential minimum during inflation,  
generically,  moduli are coherently generated after the end of inflation
and come to dominate the energy density of the universe before their decay.
The standard hot thermal universe for     
a successful Big Bang Nucleosynthesis (BBN) then requires the lightest modulus
to decay dominantly into the standard model (SM) particles well before the
BBN epoch. 
Accordingly,   the modulus has to be heavier than about 
$100$~TeV~\cite{Coughlan:1983ci, Ellis:1986zt,Goncharov:1984qm},
and its non-SM decay channels should be suppressed at least by a couple of
orders of magnitude relative to the SM ones.

As the minimal interaction of a modulus with matter fields,  
the coupling to matter kinetic terms is 
incapable of rendering the modulus to dominantly decay into the SM sector,
questioning the compatibility of string moduli with a successful cosmology.    
This is because the modulus decay into matter particles is generally suppressed by
the mass of final states, 
whereas the decay into gravitinos or stringy axions  
has no such suppression.   
The former may  cause the moduli-induced gravitino 
problem~\cite{Endo:2006zj,Nakamura:2006uc,Dine:2006ii}, 
while the latter may lead to too much dark radiation~\cite{Cicoli:2012aq,Higaki:2012ar,Higaki:2013lra}. 
A viable reheating after modulus domination thus 
necessitates specific interactions between the modulus and the SM sector.
In the Kachru-Kallosh-Linde-Trivedi (KKLT)~\cite{Kachru:2003aw} 
and LARGE volume scenarios~\cite{Balasubramanian:2005zx},  
which are the explicit realizations of four-dimensional  de Sitter vacua with all string moduli stabilized,
one may rely on modulus-dependent gauge couplings~\cite{Nakamura:2008ey} or
a modulus coupling to the Higgs bilinear operator in the K\"ahler 
potential~\cite{Cicoli:2012aq,Higaki:2012ar,Higaki:2013lra}.

In this paper we point out that  the decay of a modulus into axions,  
which is induced by the derivative interactions arising
from its coupling to the K\"ahler potential,   
can occur without suppression by the mass of final states. 
This is the case for the decay 
into stringy axions regardless of from which modulus the stringy axion
comes.
The decay into Nambu-Goldstone bosons (NGBs) also avoids mass suppression  
if the modulus is not much heavier than the scalar partner of the NGB,  i.e.~the saxion.  
All these features are  insensitive to the details of moduli stabilization and string compactification,
indicating that string moduli are basically axion-philic.
Such axion-philic nature makes  a crucial impact on cosmology because 
a sizable amount of dark radiation is produced from their decay
in string compactifications giving ultralight stringy axions,  and possibly in SM extensions
including a cosmologically stable axion,  like the QCD axion solving 
the strong CP problem~\cite{Peccei:1977hh}  or a light hidden photon.

Recovering the standard hot big bang universe for the successful BBN
requires additional couplings to the SM sector
if the universe undergoes a modulus-dominated phase.
As mentioned,   modulus-dependent gauge couplings and/or a modulus
coupling to the holomorphic bilinear operator of 
SM-charged matter  superfields in the K\"ahler potential may be necessary.
On the other hand,  if the modulus couples to a holomorphic bilinear operator 
involving the NGB superfield,  
NGBs are produced from the modulus decay 
at the rate without mass suppression. 
Also,    the fermionic partner of the NGB can contribute to dark matter
as a feebly interacting massive particle (FIMP).
If sufficiently light,  it may resolve the moduli-induced gravitino problem.

 This paper is organized as follows. 
 In section~\ref{sec:modulus-interactions},
 we explore model independent and dependent interactions of a string modulus
 inducing its decay without mass suppression,  and 
 show that  moduli are generally axion-philic insensitive to the details of model.    
 The cosmological implications of the axion-philic nature of moduli
 are discussed in section~\ref{sec:dark-radiation}.
 Section~\ref{sec:conclusions} is the conclusions.

\section{Moduli  interactions}
\label{sec:modulus-interactions} 
 
String moduli generally couple to a (pseudo) NGB and stringy axion, 
if exists,  through derivative interactions irrespective of the details of string compactification.
We examine how significantly such interactions contribute to their decay.  
We also explore model-dependent derivative interactions,  
and  discuss how to make moduli mainly decay into the SM sector 
as required for successful cosmology.

\subsection{Model-independent axionic decay modes}

It is well known that,   for a string modulus stabilized while respecting the associated 
shift symmetry, 
the decay of its real component into a pair of its imaginary one,  dubbed as {\it stringy axion},
is not mass suppressed~\cite{Cicoli:2012aq,Higaki:2012ar,Higaki:2013lra}.
It is also known that the decays of moduli into matter particles caused by
derivative interactions arising from the K\"ahler potential are however typically suppressed 
by the mass of final states. 
In this work we show that,  interestingly, such mass suppression does not occur for the decay into NGBs too
if the scalar partner of the NGB,  i.e.~the saxion, is heavier than or 
similar in mass to the modulus.
We further find that, regardless of from which modulus the stringy axion originates,
moduli decay into stringy axions without mass suppression.

In order to see the effect of derivative interactions,   
we consider a simple model with a single modulus $S$ that 
enjoys a shift symmetry,
${\rm Im}(S) \to {\rm Im}(S) + {\rm constant}$, 
at the perturbative level. 
The shift symmetry is not essential for determining the modulus decay width,
but makes our discussion simpler because 
it ensures that the modulus appears in the K\"ahler potential in the combination
of $S+S^*$.  
The K\"ahler potential is written 
\begin{equation}
K = K_0  + Z \Phi \Phi^*, 
\end{equation}
where $\Phi= \phi + \sqrt 2 \theta \psi + \theta \theta F$ represents a matter superfield,
and we set the reduced Planck mass,  $M_{Pl}=1$,  unless explicitly written. 
Here $K_0$ and $Z$ are a function of $S+S^*$.
Expanding the modulus around its vacuum expectation value (VEV)
\begin{equation}
\delta S \equiv S- S_0 
=  s  + \sqrt2 \theta \tilde s + \theta\theta F^S,
\end{equation}
one always finds the K\"ahler potential interaction of the modulus with matter superfields
\begin{equation}
\label{S-phi}
(\delta S + \delta S^*)  \Phi \Phi^*,
\end{equation}
regardless of the details of the model,  
where  the overall coupling constant is proportional to $\langle \partial_S Z \rangle$.
If matter fields do not develop a VEV, 
the modulus decay is determined simply by the interactions arising 
from the $D$-term of (\ref{S-phi}).
In terms of the component fields,  they read
\begin{equation}
\label{modulus-couplings}
\left. \delta S \Phi \Phi^* \right|_D =
s F F^*
+ \phi F^S F^*
-s \phi \partial^2 \phi^*
- i s \psi \sigma^\mu \partial_\mu \bar \psi ,
\end{equation}
omitting total derivatives. 
Written in the above form,  the derivative interactions are directly combined with
the equations of motion for $\phi$ and $\psi$ to
show that  the decay rate of the modulus is suppressed 
by the mass of decay products.

On the other hand,   if $\phi$ develops a nonzero VEV,  
$\phi$ and $\psi$ are not the mass eigenstates anymore,
and there also arises kinetic mixing between the moduli and matter fields.  
To see how the modulus decay is affected, 
we consider the case where $\phi$ spontaneously breaks a global U$(1)$ symmetry.
Then the matter scalar field is decomposed into the saxion and the NGB
\begin{equation}
\label{phi}
\phi(x) = \frac{f}{\sqrt2}\left(
1+ \frac{  \sigma(x)}{f} \right) e^{i \frac{  a(x)}{f} },
\end{equation}
with $f$ being the axion decay constant,  
while the modulus field is written
\begin{equation}
s(x) = \frac{1}{\sqrt2} \left( s_r(x) + i s_i(x) \right),
\end{equation}
for the real scalars $s_r$ and $s_i$. 
The derivative interaction from $\delta S \Phi \Phi^*$, 
\begin{equation}
\Delta {\cal L} = \sqrt 2 \kappa \left(   s \phi \partial^2 \phi^* + {\rm h.c.} \right),
\end{equation}
gives rise to the kinetic mixing terms and scalar cubic interactions 
\begin{eqnarray}
\label{kinetic-mixing}
\kappa^{-1} \Delta {\cal L}|_2 &=&
f s_r \partial^2 \sigma
+ f s_i \partial^2 a,
\\
\label{cubic-terms}
\kappa^{-1} \Delta {\cal L}|_3 &=& 
- \frac{1}{2}aa \partial^2 s_r 
+ a \sigma \partial^2 s_i  
\nonumber \\
&&
+\, ( s_r \sigma - s_i a)   \partial^2 \sigma
+ ( s_r a +  s_i \sigma  ) \partial^2 a,
\end{eqnarray}
up to total derivatives.
Here  the overall coupling constant is determined by
\begin{equation}
\label{kappa}
\kappa =-
\frac{1}{\sqrt2}
\left\langle 
( \partial^2_S   K_0 )^{-\frac{1}{2}}
 \partial_S \ln Z \right \rangle,
\end{equation} 
and it  is generally of the order unity. \footnote{
For K\"ahler moduli,   $\kappa$ is fixed by the location of the
corresponding matter field in extra dimensions.
It is then possible to suppress $\kappa$,  but up to by a loop factor because
it receives quantum corrections from non-perturbative effects,  string loops,  
and higher order $\alpha^\prime$ corrections.
}
As will be shown shortly,   the kinetic mixing has critical impacts on 
the modulus decay 
when combined with the cubic interactions arising from the matter K\"ahler potential,  
\begin{equation}
\label{matter-cubic}
\Phi\Phi^*|_{D,3} =
  \frac{1}{2f} aa \partial^2 \sigma -\frac{1}{f} \sigma a \partial^2 a,
\end{equation}
up to total derivatives.

The  mixing terms (\ref{kinetic-mixing}) should be removed  to
examine  how much  the derivative interactions (\ref{cubic-terms}) contribute 
to the modulus decay.
The kinetic mixing between the CP-odd scalar bosons is eliminated by taking
the field transformation
\begin{equation}
s_i \to s_i,  \quad
a \to a + \kappa f s_i.
\end{equation}
After the field transformation,  i.e.  in the canonical  mass basis,  
all  the derivative cubic couplings of $s_i$ vanish because the contributions  from (\ref{cubic-terms})
and (\ref{matter-cubic}) exactly cancel each other.   
Hence there remain no interactions inducing the decay of $s_i$.  
For the CP-even scalar bosons,  one can remove their kinetic mixing by taking the field transformation 
\begin{equation}
s_r \to  s_r 
+ \epsilon_1 \kappa f \sigma,
\quad
\sigma \to \sigma
+\epsilon_2  \kappa f s_r,
\end{equation} 
for $f\ll  M_{Pl}$,  with the coefficients  determined by 
\begin{equation}
\label{coefficients}
\epsilon_1  \simeq -\frac{m^2_\sigma}{m^2_{s_r} - m^2_\sigma},
\quad
\epsilon_2  \simeq  \frac{m^2_{s_r} }{m^2_{s_r} - m^2_\sigma},
\end{equation}
where $m_i$  denotes  the mass of the indicated scalar boson,
and we have neglected small corrections suppressed by $(f/M_{Pl})^2$.
As a result,  from (\ref{cubic-terms}) and (\ref{matter-cubic}), 
the derivative interactions for $s_r$ read
\begin{equation}
\Delta {\cal L}|_3
= \kappa s_r \sigma \partial^2 \sigma
+ \frac{\kappa}{2} \frac{m^2_\sigma}{m^2_{s_r} - m^2_\sigma}
aa \partial^2 s_r,
\end{equation}
in the canonical mass basis. 
Therefore,  if $m_{s_r} \gg m_\sigma$, 
the equations of motion lead to that both $s_r \to aa$ and
$s_r \to \sigma \sigma$ are suppressed by the saxion mass.  
On the other hand,  
if $m_{s_r} \ll m_\sigma$, i.e.~if the saxion is heavier than the modulus,
the relevant interaction for the modulus decay becomes
\footnote{ 
Eq.~(\ref{coefficients}) is valid for  $f/M_{Pl} \ll |m^2_{s_r} - m^2_\sigma|/(m^2_{s_r} + m^2_\sigma)$.
Although it is unlikely,  if the masses of the modulus and saxion are close to each other,
the coupling of $aa\partial^2 s_r$ induced by the kinetic mixing can be much larger than $\kappa$.
Such enhancement can make the coupling of $s_r$ to NGBs suppressed not by $M_{Pl}$ but by $f$ around
the maximal mixing.   
}
\begin{equation}
\Delta {\cal L}|_3 \ni
-\frac{\kappa}{2} aa \partial^2 s_r.
\end{equation}   
The above shows that the modulus gets axion-philic,  and decays
into NGBs at the rate
\begin{equation}
\Gamma(s_r \to aa) = \frac{\kappa^2}{32\pi} \frac{m^3_{s_r}}{M^2_{Pl}},
\end{equation} 
where the masses of final states have been neglected. 
It should be noted that non-suppression of the axionic decay mode  $s_r \to aa$,
which holds as long as
the modulus is lighter than or similar in mass to the scalar partner of the NGB,
is a model-independent generic feature of string moduli.

The non-derivative interactions involving $F$ and $F^S$ in (\ref{modulus-couplings})
also mediate the decay, 
which is however suppressed by the mass of final states or further. 
This can be understood as follows. 
The matter $F$-component is expanded in powers of $\phi$ as  
\begin{equation}
\label{F-term}
F = \langle F \rangle + \langle \partial_\phi F \rangle \phi  
+ \langle \partial_{\phi^*} F \rangle \phi^*
+ \cdots,
\end{equation}
while the scalar mass,  $m_\phi$, is determined by
\begin{equation}
m^2_\phi = \left \langle \frac{\partial^2 V}{\partial \phi \partial \phi^*} \right\rangle
=
|\langle \partial_\phi F \rangle|^2 + |\langle \partial_{\phi^*} F \rangle|^2
+ \cdots,
\end{equation}
where $V$ is the full scalar potential,
and the ellipsis denotes possible extra contributions. 
Hence,  the coefficients of the linear terms in (\ref{F-term}) are bounded by
\begin{equation}
|\langle \partial_\phi F \rangle|,\,
|\langle \partial_{\phi^*} F \rangle|   \lesssim m_\phi, 
\end{equation}
barring cancellation among various contributions to the scalar mass. 
Expanded in terms of $s$ and $s^*$,  the modulus $F$-component 
exhibits the similar feature.

Let us move on to stringy axions that appear in string compactification such that 
a modulus,  say $T$,  is stabilized while preserving the associated shift symmetry. 
The interactions responsible for the modulus decay come from the K\"ahler potential
\begin{equation}
(\delta S + \delta S^*) (\delta T + \delta T^*)^2,
\end{equation}
where $\delta S$ and $\delta T$ are moduli fluctuations around the vacuum.
The $D$-term of the holomorphic cubic term,  $\delta S \delta T \delta T$,
includes derivative interactions, which are however summed to be a total derivative. 
The relevant derivative interactions from the above K\"ahler potential
term read
\begin{equation}
\Delta {\cal L}|_3 = 
\sqrt 2 \hat \kappa 
\left(\frac{1}{2}
t t\partial^2 s^* + s t \partial^2 t^* \right)
+ {\rm h.c.}, 
\end{equation}
up to total derivatives.
The coupling constant is determined by
\begin{equation}
\hat \kappa
= -\frac{1}{\sqrt2}
\left\langle
(\partial^2_S K_0)^{-1/2}(\partial^2_T K_0)^{-1}
\partial_S\partial^2_T K_0
\right\rangle,
\end{equation}
which is of the order unity in general.   
Here  $t$ is the scalar component of $\delta T$ and is decomposed as
\begin{equation}
t = \frac{1}{\sqrt2}(\tau + i \varphi), 
\end{equation}
for real scalars $\tau$ and $\varphi$.
The stringy axion,  $\varphi$,  remains massless until one adds non-perturbative effects 
breaking the shift symmetry explicitly.   
It is easy to see that the derivative interactions are written in terms of the real scalar fields
as 
\begin{eqnarray} 
\hat \kappa^{-1} \Delta {\cal L}|_3 &=&
\frac{1}{2} \tau\tau \partial^2 s_r
- \frac{1}{2} \varphi\varphi \partial^2 s_r
+  \varphi \tau \partial^2 s_i
\nonumber \\
&&
+\,
 (\tau s_r - \varphi s_i)\partial^2 \tau
+  ( \tau s_i + \varphi s_r) \partial^2 \varphi.
\end{eqnarray}  
The first three interactions are expected to give a significant contribution to the decays 
of $s_r$ and $s_i$,
while the decay via the others is mass suppressed.

To estimate correctly how fast the modulus decays via the derivative interactions,  
we need to remove kinetic mixing induced by the K\"ahler potential 
$(\delta S + \delta S^*) (\delta T + \delta T^*)$.
Combined with the derivative interactions from 
$(\delta S+ \delta S^*)^3$,   $(\delta S + \delta S^*)^2 (\delta T + \delta T^*)$
and  $(\delta T + \delta T^*)^3$,
the kinetic mixing modifies the couplings of the moduli interactions involving $\partial^2 s_r$ and
$\partial^2 s_i$. 
However,  their values in the canonical mass basis generally remain the same order of
magnitude  because  the kinetic mixing is independent of 
$\hat \kappa$,   differently from the NGB case where 
 the interactions (\ref{kinetic-mixing}) and (\ref{cubic-terms})
are all originated from the same single K\"ahler potential term.   
The modulus decay into stringy axions is therefore given by
\begin{equation}
\Gamma(s_r \to \varphi\varphi )
=  \frac{\kappa^{\prime 2} }{32\pi} \frac{m^3_{s_r}}{M^2_{Pl}},
\end{equation}
where $\kappa^\prime$, 
which depends on $\hat \kappa$ and  kinetic mixing between the moduli,
 is generally of the order unity. 
Note that this
non-suppression of modulus decay into stringy axions is a quite general feature of string
compactification,
and holds   regardless of which one of $s_r$ and $\tau$
is heavier.

The importance of modulus decay to stringy axions has been noticed
for the case where $S$ and $T$ are identical,  
in the the LARGE volume scenario~\cite{Cicoli:2012aq,Higaki:2012ar},
and later in the generalized setup~\cite{Higaki:2013lra}.
It is worth notifying that,   if string compactification involves a very light stringy axion,
not only its scalar partner but also any other moduli decay into stringy axions 
with a sizable branching fraction insensitively to the details of moduli stabilzation.  
For instance,  a massless stringy axion arises in the generalized KKLT with multiple 
K\"ahler moduli if a modulus,  which does not appear in the superpotential,  
is stabilized by K\"ahler mixing with the others~\cite{Choi:2006za}. 
In such a case, 
the decay into stringy axions is sizable for all the real components of moduli.

\subsection{Model-dependent decay modes}
\label{subsec:model-dependent}

In this subsection we explore model-dependent interactions considerably
contributing to the modulus decay,  
and discuss how to make the modulus mainly decay into the SM sector.   
Let us begin with a modulus coupling to the holomorphic bilinear operator
of matter superfields 
\begin{equation}
\label{K-bilinear} 
 (\delta S + \delta S^*) 
(\Phi_1 \Phi_2 + {\rm h.c.}),
\end{equation} 
in the K\"ahler potential. 
It is the derivative interactions including only the scalar fields that
can enhance the modulus decay.
Whereas the holomorphic cubic term,  $\delta S \Phi_1 \Phi_2$, 
is irrelevant because its $D$-term only gives  a total derivative,
the  non-holomorphic cubic term gives
\begin{equation}
\label{bilinear-term}
\delta S^* \Phi_1 \Phi_2|_D 
= -\phi_1\phi_2 \partial^2 s^*,
\end{equation}
up to total derivatives.
It is obvious that the decay of the modulus via the interaction  (\ref{bilinear-term})
is not suppressed by the mass of final states.    
Hence,  an interaction of the type  (\ref{K-bilinear})  can enhance 
the modulus decay into the SM sector 
if the involved matter fields are SM-charged.

More concretely,  one can consider a modulus coupling to $QQ^c$ in the K\"ahler 
potential with a coupling constant of the order unity or larger.
Here the matter superfields $Q+Q^c$ are vector-like under the SM gauge groups, 
and are lighter than the modulus. 
A natural candidate of such operators is 
the Higgs bilinear $H_uH_d$~\cite{Cicoli:2012aq,Higaki:2012ar,Higaki:2013lra},
where $H_u$ and $H_d$ are respectively the up- and down-type Higgs doublet.
The modulus coupling to $H_uH_d$ in the K\"ahler potential
\begin{equation}
\label{HuHd1}
\Delta K_1 =  \xi H_u H_d + {\rm h.c.}
\end{equation}
includes the term
\begin{equation}
\langle \partial_S \xi \rangle (\delta S +\delta S^*) H_u H_d + {\rm h.c.},
\end{equation}
where $\xi$ is a function of $S+S^*$. 
Meanwhile, it gives a contribution to the Higgsino mass parameter $\mu$
and the Higgs mixing parameter $B$ as
\begin{equation}
\Delta \mu = \langle \xi \rangle m_{3/2},
\quad \Delta B = -m_{3/2},
\end{equation}
with $m_{3/2}$ being the gravitino mass,  
because it explicitly breaks the super-Weyl symmetry.
It is then difficult to avoid the Higgs $\mu/B\mu$ problem 
in the scenario where the gravitino is much heavier than the visible sparticles,  
i.e.~if anomaly mediation,   which is  a model-independent source of 
soft supersymmetry breaking masses
in supergravity,  is sizable
as is the case in the KKLT moduli stabilization. 
To allow an order unity coupling of $\delta S$ to $H_uH_d$
while suppressing the contributions to $\mu$ and $B$, 
one can consider
\begin{equation}
\label{HuHd2}
\Delta K_2 = \xi \frac{X^*}{X} H_u H_d + {\rm h.c.},
\end{equation}
for a singlet $X$ that is radiatively stabilized
by the loop potential~\cite{Nakamura:2008ey}.
As induced by the above super-Weyl invariant K\"ahler potential term, 
both $\Delta \mu$ and $\Delta B$ are loop suppressed relative to $m_{3/2}$
as required for the correct electroweak symmetry breaking, 
while the modulus coupling to $H_uH_d$ is still given by $\langle \partial_S \xi \rangle$
and is generally of the order unity.

Let us examine how the modulus decay is affected by the interaction $(\ref{K-bilinear})$
when both $\phi_1$ and $\phi_2$ develop a VEV to spontaneously break the
global U$(1)$ symmetry,  under which $\phi_1$ and $\phi_2$ carry charge
$1$ and $-1$,  respectively, so that $\phi_1\phi_2$ becomes U$(1)$ invariant. 
The scalar fields are decomposed as
\begin{equation}
\phi_i = \frac{f_i}{\sqrt2}
\left( 1 + \frac{\sigma_i}{f_i} \right) e^{i \frac{a_i}{f_i} },
\end{equation}
for $i=1,2$.
The massless NGB is then written  
\begin{equation}
a = \frac{f_1 a_1 - f_2 a_2}{\sqrt{f^2_1 + f^2_2}}. 
\end{equation} 
Through the interaction (\ref{bilinear-term}), the modulus couples 
to the two massive scalar bosons composed of $\sigma_i$ and 
the other combination of $a_i$ proportional to $f_2 a_1 + f_1 a_2$,
but it is forbidden to couple to the NGB due to the U$(1)$ symmetry. 
Nonetheless, in the presence of the coupling  (\ref{bilinear-term}), 
NGBs can be produced abundantly from the cascade decay of moduli
because the saxion generally decays mainly into NGBs via the $aa\partial^2 \sigma$ interaction.

Another role of the interaction (\ref{bilinear-term}) is to induce kinetic mixing between 
the modulus and $\sigma_i$ after spontaneous U$(1)$ breaking.
Combined with the interactions $a_ia_i \partial^2 \sigma_i$ 
coming from the kinetic term of $\phi_i$,
such kinetic mixing generates the $a a \partial^2 s_r$ interaction whose coupling is given by  
\begin{equation}
\frac{2f_1 f_2}{f^2_1 + f^2_2} 
\end{equation}
times the coupling of (\ref{bilinear-term}),
for  the case where
the modulus is much heavier than the massive scalar bosons composed of $\sigma_i$.
This implies that,   after  electroweak symmetry breaking, 
the modulus coupling to $H_uH_d$ in the K\"ahler potential induces the modulus decay
into the CP even and odd neutral Higgs bosons,  and the charged Higgs bosons
without mass suppression. 
In addition,  due to the kinetic mixing,   it contributes to the modulus decay into the associated NGBs,
i.e.~into the longitudinal modes of the weak gauge bosons,
according to  the Goldstone boson equivalence theorem~\cite{Cornwall:1974km,Vayonakis:1976vz}

As a model-dependent modulus decay mode,  one can also consider 
the decay into the  gauge sector,  which
  is possible if the modulus appears in the gauge kinetic function
\begin{equation}
f_a = S + \Delta f_a,
\end{equation}
with $\Delta f_a$ being a $S$-independent constant.   
It is straightforward to see that the decay rate into gauge bosons reads~\cite{Nakamura:2008ey}  
\begin{equation}
\Gamma(s_r \to gg) = \frac{N_g}{128\pi} \kappa^2_g \frac{m^3_{s_r}}{M^2_{Pl}},
\end{equation}
where $N_g=12$ counts the number of gauge bosons, 
and the order unity constant $\kappa_g$ is determined by
\begin{equation}
\kappa_g = \left\langle (\partial^2_S K_0)^{-\frac{1}{2}}  
\frac{1}{{\rm Re} f_a }  \right\rangle.
\end{equation}
On the other hand,   the decay rate into gauginos is written
\begin{equation}
\Gamma(s_r \to \tilde g \tilde g) = 
\beta^2_r \Gamma(s_r \to gg),
\end{equation}
with $\beta_r$ defined by
\begin{equation} 
\langle \partial_s F^S \rangle s
+ \langle \partial_{s^*} F^S \rangle s^*
\equiv \beta_r m_{s_r} s_r
+ \beta_i m_{s_i} s_i,
\end{equation}
where $\beta_r$ and $\beta_i$ do not exceed order unity in size. 
Here the decay rates have been evaluated neglecting the mass of decay products. 
The decay rates of $s_i$ can be read off from those for $s_r$ by taking the replacements,
$m_{s_r}\to m_{s_i}$ and $\beta_r\to \beta_i$.
Note that the KKLT leads to $\beta_r\simeq \beta_i \simeq 1$ for $\Delta f_a=0$.  
In order to enhance the decay into the visible sector further,  one may consider 
the case where  the Higgsino mass parameter is generated from the K\"ahler 
potential 
as (\ref{HuHd1})~\cite{Cicoli:2012aq,Higaki:2012ar}
or as (\ref{HuHd2}).

We close this section by mentioning that the modulus can 
decay to a gravitino pair at a sizable rate if kinematically allowed.
In particular,  the decay is dominated by the coupling to the helicity $\pm 1/2$ components,  
i.e.~to the Goldstino,
which is  proportional to the modulus 
$F$-term~\cite{Endo:2006zj,Nakamura:2006uc,Dine:2006ii}.
The decay rate into gravitinos reads
\begin{equation}
\Gamma(s_r \to \tilde G \tilde G ) = 
\frac{\kappa^2_{3/2}}{288\pi} \frac{m^3_{s_r}}{M^2_{Pl}},
\end{equation}
in the limit $m_{3/2}\ll m_{s_r}$, 
with $m_{3/2}$ being the gravitino mass. 
Here $\kappa_{3/2}$ is defined as follows
\begin{equation}
\kappa_{3/2} \equiv
\left \langle (\partial^2_S K_0)^{\frac{1}{2}} F^S \right \rangle
\frac{m_{s_r}}{m^2_{3/2}},
\end{equation} 
whose size is below order unity because the modulus interactions
are gravitational.  
If it  is of the order unity,  as is typically the case  in 
the known examples of moduli stabilization,   
a large gravitino yield after moduli decay can cause  cosmological 
difficulties,
dubbed as the moduli-induced gravitino problem.

\section{Cosmological implications}
\label{sec:dark-radiation}

Reheating takes place via  moduli decay if 
the universe passes through a modulus-dominated epoch 
as  generally expected in string compactifications.
To catch the relevant features of such driven reheating,  we simply consider  
a single modulus $S$
under the assumption that its real and imaginary component have 
a similar mass,  $m_s$.\footnote{
For the case of multiple moduli,  
the decay properties discussed in section~\ref{sec:modulus-interactions} apply to all the moduli. 
If not much heavier than the lightest one, 
heavy moduli can still be cosmologically important. 
For instance,   if there were two moduli,   $S_1$ and  $S_2$, 
with masses $m_1$ and $m_2$ ($m_1<m_2$),   respectively, 
the energy density of the decay products of $S_2$  is diluted by 
the entropy released from late-time decay of $S_1$.
The dilution factor is given by the ratio between their decay temperatures,   and 
reads
$\Delta \sim (m_2/m_1)^{3/2}$,
assuming that the displacement of moduli at the onset of  coherent oscillation
is of the order of the  Planck scale.  
}
In flux compactifications,  a modulus unfixed by flux can be stabilized by non-perturbative dynamics 
or K\"ahler mixing with others.
Its mass is then tied to the scale of supersymmetry breaking,
$m_s \sim m_{3/2}$,  up to a factor of order $\ln(M_{Pl}/m_{3/2})$~\cite{Choi:2005ge,Choi:2008hn}. 
On the other hand,  in the LARGE volume scenario,  
 the mass of the large volume modulus  can be much smaller 
 than $m_{3/2}$~\cite{Balasubramanian:2005zx}.

To allow successful BBN after the modulus dominated phase,  
the modulus should decay dominantly into the visible sector while forming 
the standard thermal background with temperature $T_s$ constrained 
by~\cite{deSalas:2015glj,Hasegawa:2019jsa}  
\begin{equation}
T_s \simeq  \left( \frac{90}{\pi^2 g_\ast} \right)^{\frac{1}{4} }
( \Gamma_s M_{Pl})^{\frac{1}{2}} 
\gtrsim 5\,{\rm MeV},
\end{equation}
where $g_\ast$ is the relativistic degrees of freedom at $T_s$, 
and $\Gamma_s$ is the total decay width of the modulus.
Here we have assumed an instantaneous conversion of the modulus energy 
density to radiation.
The  constraint on $T_s$ requires the modulus to be heavier than about $100$~TeV.
Modulus domination also constrains models for
baryon and dark matter genesis
because  a huge amount of entropy is released from moduli decay.
If the genesis occurs at the early stage of moduli domination or before,   
the dilution factor  naively reads,
$\Delta \sim 10^{12}\,(m_s/{\rm PeV})^{-1}$,  since 
the initial displacement of the modulus after inflation is generally of the order
of the Planck scale.

A viable reheating requires the modulus to decay dominantly to the visible sector.
This can be achieved by adding model-dependent modulus couplings
such as  a coupling to the bilinear operator $QQ^c$ or $H_uH_d$ in the K\"ahler potential,
or to the visible gauge sector 
as discussed in subsection~\ref{subsec:model-dependent}.
In the presence of such couplings,  the decay rate into the visible sector reads
\begin{equation}
\Gamma_{s\to {\rm SM}} 
= \frac{\kappa^2_{\rm SM}}{32\pi}
\frac{m^3_s}{M^2_{Pl}},
\end{equation}
where $\kappa_{\rm SM}$ is a constant of order unity.   
As discussed already,
the modulus generally decays  into axions without mass suppression if
string compactification involves a light stringy axion,  $\varphi$, 
or if there is a (pseudo) NGB,  $a$,  whose scalar partner has a mass larger than or comparable
to the modulus mass. 
A theoretically well-motivated NGB is the QCD axion~\cite{Ringwald:2012hr}. 
The branching ratio of the axion-philic modulus into axions is written
\begin{equation}
{\rm Br}(s\to {\rm axions}) 
= \frac{\kappa^2 + \kappa^{\prime 2}}{\kappa^2_{\rm SM}},
\end{equation}
which naturally lies in the range of the order of $0.1$. 
If the axions are cosmologically stable,  the produced axions form dark radiation.
The contribution to the effective number of neutrino species is  estimated by~\cite{Choi:1996vz} 
\begin{equation}
\Delta N_{\rm eff}  
= \frac{43}{7}
\frac{{\rm Br}(s\to {\rm axions}) }{1-{\rm Br}(s\to {\rm axions}) }
\left( \frac{g_\ast(T_{\nu,{\rm dec}})  }{g_\ast(T_{s})} \right)^{\frac{1}{3}},
\end{equation}
under the assumption that the branching ratio into the visible sector is given by 
$1-{\rm Br}(s\to {\rm axions})$.
Here $T_{\nu,{\rm dec}}$ is the neutrino decoupling temperature. 
An observable amount of dark radiation is a natural prediction of the axion-philic modulus
insensitively to the detail of moduli stabilization.    
Dark radiation can have interesting cosmological effects in the early universe.
For instance,  if $\Delta N_{\rm eff}\gtrsim 0.5$,  which is the case
for ${\rm Br}(s\to {\rm axions})$ larger than about  $0.07$,  
it may relieve the Hubble tension~\cite{Riess:2019cxk,DiValentino:2021izs}.

It should be noted that the presence of a plenitude of light stringy axions,  
an axiverse,  has been suggested as evidence for the extra dimensions of 
string theory~\cite{Arvanitaki:2009fg}.
In the axiverse,  the lightest modulus would dominantly decay  into these axions, 
producing too much dark radiation if the universe undergoes a modulus-dominated phase.  
The modulus coupling to $H_uH_d$ or to the gauge sector would not be sufficient
to suppress ${\rm Br}(s\to {\rm axions})$.
To avoid the moduli-induced axion problem, 
i.e.~to enhance the modulus decay into the visible sector, 
one may thus rely on the modulus coupling to $QQ^c$ in the K\"ahler potential,
for instance,   for  a number of  $Q+Q^c$ that form complete SU$(5)$ multiplets.
Here the supersymmetric mass of $Q+Q^c$ in the superpotential is assumed
to be smaller than the modulus mass. 
We also note  that,  depending on their masses,  
the number of $Q+Q^c$ is constrained by the perturbativity of the gauge interactions
up to the grand unification scale.   
The phenomenological and cosmological details of such particles are out of
the scope of this paper,  and hence will not be touched here.

The cosmological features discussed so far remain the same 
even if the saxion has an initial amplitude of the order of the Planck scale,
as long as
it decays much faster than the modulus,  i.e.~if
\begin{equation}
m_\sigma \gtrsim 100\,{\rm GeV} \left( \frac{f}{10^{12}{\rm GeV}} \right)^{\frac{2}{3}}
\left(\frac{m_s}{\rm PeV} \right).
\end{equation}
However,  the situation changes much  if the saxion is thermally trapped at the origin to drive
thermal inflation~\cite{Lyth:1995hj}.
It is also a plausible scenario although we do not consider it in this paper.

Moduli stabilization with $m_s > 2 m_{3/2}$ generally confronts 
the moduli-induced gravitino problem unless one adds $R$-parity violating interactions.  
Interestingly,   the gravitino problem can be relaxed  if the fermionic partner of the axion,   the axino,
is much lighter than the lightest observable sparticle (LOSP) since otherwise
LOSPs produced from gravitino decay would overclose the universe 
under the $R$-parity conservation.
Even for moduli stabilization with $m_s< 2m_{3/2}$,  the axino is still an attractive particle 
because, as a FIMP, 
it is a viable alternative to weakly interacting massive particle (WIMP) cold dark matter.

In the moduli-dominated universe, 
axinos are non-thermally produced from the decays of the modulus,   gravitino, 
and LOSP.
The axino abundance from LOSPs decay is quite model-dependent because 
LOSPs can efficiently annihilate before their  decay depending on the LOSP nature
and the value of $f$.  
Meanwhile,   independently of model details,  
the axino yield directly produced from the gravitationally interacting moduli and gravitinos
is estimated by
\begin{equation}
Y_{\tilde a}|_{s\to \tilde a} =2 B^s_{\tilde a} \frac{3}{4}  
\frac{T_{s}}{m_s},
\end{equation}
where $B^s_{\tilde a}$ is defined by
\begin{equation}
B^s_{\tilde a} \equiv {\rm Br}(s\to \tilde a\tilde a)  + {\rm Br}(s\to \tilde G\tilde G){\rm Br}(\tilde G \to \tilde a),
\end{equation}
with  ${\rm Br}(s\to \tilde a\tilde a)\sim (m_{\tilde a}/m_s)^2$ as follows from
(\ref{modulus-couplings}).
Thus,  in the case with $m_s<2m_{3/2}$,  the axino should have mass, 
\begin{equation}
m_{\tilde a} \lesssim 1\,{\rm TeV}
\left( \frac{0.1{\rm GeV}}{T_{s}} \right)^{\frac{1}{3}}
\left( \frac{m_s}{\rm PeV} \right).
\end{equation}
not to exceed the observed dark matter density. 
On the other hand,  if $m_s>2m_{3/2}$,  the gravitino produced moduli 
uniformly decays to the axino and all lighter visible sparticles,  
and should be heavier than a few tens of TeV in order to decay before  BBN~\cite{Kawasaki:2017bqm}.
The branching fraction of the modulus decay to gravitinos  is generally sizable,  indicating
that the axino should  be much lighter than the above bound. 
The axino mass and the LOSP nature may further be constrained 
by the axino free-streaming scale at  the time of
matter-radiation equality,   which should be less than the order of Mpc to be 
consistent with the observations of Lyman-$\alpha$ forest~\cite{Viel:2005qj}.
Axinos are mostly produced from the decay of LOSPs 
as long as the number of the visible sparticles lighter than the gravitino
is much larger than unity.  
Then,  the free-streaming constraint is relaxed 
if the LOSP scattering rate is larger than its decay rate so that
LOSPs become non-relativistic before their decay.   
See  refs.~\cite{Nakamura:2008ey,Bae:2021rmg} 
for some examples in the KKLT scenario.  
Finally,   we note that axinos are also produced via freeze-in processes,
which can be effective depending of the value of $f$ if 
$T_s$ is not much lower than the freeze-out temperature
of the LOSP~\cite{Bae:2021rmg}.

\section{Conclusions}
\label{sec:conclusions}

In this work we have showed that string moduli have  axion-philic nature.
Induced by the model-insensitive derivative interactions,
the decay of a modulus into stringy axions occurs without suppression by the mass of decay products.  
Interestingly,  we found that 
this feature holds not only for the scalar partner of the stringy axion but also for any other moduli.
The decay into (pseudo) NGBs also avoids mass suppression if the modulus is not much heavier than the saxion.  
The axion-philic nature makes string moduli a natural source of an observable amount of dark radiation
in the universe that passes through a modulus-dominated era 
as is  generally expected in scenarios of  string compactification.
Proper reheating would then require moduli to have additional couplings to enhance their decay into 
the SM sector so that hot thermal bath composed of mostly  SM particles is formed. 
For instance,  the modulus can couple to the gauge sector if it appears in the gauge kinetic function.
Another way is to add  a modulus coupling to the holomorphic bilinear operator  of matter superfields in the K\"ahler 
potential.
It is also interesting to note that the fermionic superpartner of a NGB as well as coherently oscillating NGBs
can be a viable alternative to WIMP dark matter. 
\\

\noindent{\bf Acknowledgments}
 
The authors thank to Fuminobu Takahashi for helpful discussion.   
This work was supported by  the National Research Foundation of Korea (NRF) grant funded 
by the Korea government: 2018R1C1B6006061,  2021R1A4A5031460 (K.S.J.), 
2017R1D1A1B06035959 (W.I.P.),  and also by Research Base 
Construction Fund Support Program funded by Jeonbuk National University in 2021 (W.I.P.).   
%



\begin{thebibliography}{99}
 
\bibitem{Coughlan:1983ci}
G.~D.~Coughlan, W.~Fischler, E.~W.~Kolb, S.~Raby and G.~G.~Ross,
Phys. Lett. B \textbf{131} (1983), 59-64.

\bibitem{Ellis:1986zt}
J.~R.~Ellis, D.~V.~Nanopoulos and M.~Quiros,
Phys. Lett. B \textbf{174} (1986), 176-182.

\bibitem{Goncharov:1984qm}
A.~S.~Goncharov, A.~D.~Linde and M.~I.~Vysotsky,
Phys. Lett. B \textbf{147} (1984), 279-283.



\bibitem{Endo:2006zj}
M.~Endo, K.~Hamaguchi and F.~Takahashi,
Phys. Rev. Lett. \textbf{96} (2006), 211301
[arXiv:hep-ph/0602061 [hep-ph]].

\bibitem{Nakamura:2006uc}
S.~Nakamura and M.~Yamaguchi,
Phys. Lett. B \textbf{638} (2006), 389-395
[arXiv:hep-ph/0602081 [hep-ph]].


\bibitem{Dine:2006ii} 
  M.~Dine, R.~Kitano, A.~Morisse and Y.~Shirman,
  Phys.\ Rev.\ D {\bf 73}, 123518 (2006)
  [hep-ph/0604140].
  

\bibitem{Cicoli:2012aq}
M.~Cicoli, J.~P.~Conlon and F.~Quevedo,
Phys. Rev. D \textbf{87} (2013) no.4, 043520
[arXiv:1208.3562 [hep-ph]].


\bibitem{Higaki:2012ar}
T.~Higaki and F.~Takahashi,
JHEP \textbf{11} (2012), 125
[arXiv:1208.3563 [hep-ph]].


\bibitem{Higaki:2013lra}
T.~Higaki, K.~Nakayama and F.~Takahashi,
JHEP \textbf{07}, 005 (2013)
[arXiv:1304.7987 [hep-ph]].


\bibitem{Kachru:2003aw}
S.~Kachru, R.~Kallosh, A.~D.~Linde and S.~P.~Trivedi,
Phys. Rev. D \textbf{68} (2003), 046005
[arXiv:hep-th/0301240 [hep-th]].


\bibitem{Balasubramanian:2005zx}
V.~Balasubramanian, P.~Berglund, J.~P.~Conlon and F.~Quevedo,
JHEP \textbf{03} (2005), 007
[arXiv:hep-th/0502058 [hep-th]].

 

\bibitem{Nakamura:2008ey}
S.~Nakamura, K.~i.~Okumura and M.~Yamaguchi,
Phys. Rev. D \textbf{77} (2008), 115027
[arXiv:0803.3725 [hep-ph]].










\bibitem{Peccei:1977hh} 
  R.~D.~Peccei and H.~R.~Quinn,
  Phys.\ Rev.\ Lett.\  {\bf 38}, 1440 (1977);
%
  Phys.\ Rev.\ D {\bf 16}, 1791 (1977).


  
\bibitem{Choi:2006za}
K.~Choi and K.~S.~Jeong,
JHEP \textbf{01} (2007), 103
[arXiv:hep-th/0611279 [hep-th]].  
    
    
  
    

\bibitem{Cornwall:1974km}
J.~M.~Cornwall, D.~N.~Levin and G.~Tiktopoulos,
Phys. Rev. D \textbf{10} (1974), 1145
[erratum: Phys. Rev. D \textbf{11} (1975), 972].

\bibitem{Vayonakis:1976vz}
C.~E.~Vayonakis,
Lett. Nuovo Cim. \textbf{17} (1976), 383.
    
    
  
\bibitem{Choi:2005ge}
K.~Choi, A.~Falkowski, H.~P.~Nilles and M.~Olechowski,
Nucl. Phys. B \textbf{718} (2005), 113-133
[arXiv:hep-th/0503216 [hep-th]].  
  
\bibitem{Choi:2008hn}
K.~Choi, K.~S.~Jeong and K.~I.~Okumura,
JHEP \textbf{07} (2008), 047
[arXiv:0804.4283 [hep-ph]].  
    
  
  
\bibitem{deSalas:2015glj}
P.~F.~de Salas, M.~Lattanzi, G.~Mangano, G.~Miele, S.~Pastor and O.~Pisanti,
Phys. Rev. D \textbf{92} (2015) no.12, 123534
[arXiv:1511.00672 [astro-ph.CO]].



\bibitem{Hasegawa:2019jsa}
T.~Hasegawa, N.~Hiroshima, K.~Kohri, R.~S.~L.~Hansen, T.~Tram and S.~Hannestad,
JCAP \textbf{12} (2019), 012
[arXiv:1908.10189 [hep-ph]].
  
  

  
  
\bibitem{Ringwald:2012hr}
For recent reviews,  see,  for instance, 
A.~Ringwald,
Phys. Dark Univ. \textbf{1} (2012), 116-135
[arXiv:1210.5081 [hep-ph]];
M.~Kawasaki and K.~Nakayama,
Ann. Rev. Nucl. Part. Sci. \textbf{63} (2013), 69-95
[arXiv:1301.1123 [hep-ph]].  
  
  
\bibitem{Choi:1996vz}
K.~Choi, E.~J.~Chun and J.~E.~Kim,
Phys. Lett. B \textbf{403} (1997), 209-217
[arXiv:hep-ph/9608222 [hep-ph]].  


  
  
\bibitem{Riess:2019cxk}
A.~G.~Riess, S.~Casertano, W.~Yuan, L.~M.~Macri and D.~Scolnic,
Astrophys. J. \textbf{876} (2019) no.1, 85
[arXiv:1903.07603 [astro-ph.CO]].  
  
 
\bibitem{DiValentino:2021izs}
E.~Di Valentino, O.~Mena, S.~Pan, L.~Visinelli, W.~Yang, A.~Melchiorri, D.~F.~Mota, A.~G.~Riess and J.~Silk,
Class. Quant. Grav. \textbf{38} (2021) no.15, 153001
[arXiv:2103.01183 [astro-ph.CO]].
  
 
\bibitem{Arvanitaki:2009fg}
A.~Arvanitaki, S.~Dimopoulos, S.~Dubovsky, N.~Kaloper and J.~March-Russell,
Phys. Rev. D \textbf{81} (2010), 123530
[arXiv:0905.4720 [hep-th]]. 
 
  
\bibitem{Lyth:1995hj}
D.~H.~Lyth and E.~D.~Stewart,
Phys. Rev. Lett. \textbf{75} (1995), 201-204
[arXiv:hep-ph/9502417 [hep-ph]];
Phys. Rev. D \textbf{53} (1996), 1784-1798
[arXiv:hep-ph/9510204 [hep-ph]].


\bibitem{Kawasaki:2017bqm}
M.~Kawasaki, K.~Kohri, T.~Moroi and Y.~Takaesu,
Phys. Rev. D \textbf{97} (2018) no.2, 023502
[arXiv:1709.01211 [hep-ph]].



\bibitem{Viel:2005qj}
M.~Viel, J.~Lesgourgues, M.~G.~Haehnelt, S.~Matarrese and A.~Riotto,
Phys. Rev. D \textbf{71} (2005), 063534
[arXiv:astro-ph/0501562 [astro-ph]].


\bibitem{Bae:2021rmg}
K.~J.~Bae and K.~S.~Jeong,
Phys. Rev. D \textbf{104} (2021) no.1, 015013
[arXiv:2105.08236 [hep-ph]].





  \end{thebibliography}
 \end{document}